\documentclass[preprint,12pt,authoryear]{elsarticle}

\usepackage{amssymb}
\usepackage{amsmath}
\usepackage{mathtools}
\usepackage{bm}
\usepackage{xcolor}

\newcommand{\CV}{\mathcal{V}}
\newcommand{\kl}[2]{D_{\mathrm{KL}}(#1~\|~#2)}

\journal{International Journal of Multiphase Flow}

\begin{document}

\begin{frontmatter}



\title{Generative diffusion models for synthetic trajectories of heavy and light particles in turbulence} %% Article title


%\author{} %% Author name

% \affiliation{organization={},%Department and Organization

\author[label1]{Tianyi Li}
\author[label1]{Samuele Tommasi}
\author[label1]{Michele Buzzicotti}
\author[label1]{Fabio Bonaccorso}
\author[label1]{Luca Biferale}

\affiliation[label1]{organization={Department of Physics and INFN, University of Rome ``Tor Vergata''},%Department and Organization
            addressline={Via della Ricerca Scientifica 1}, 
            city={Rome},
            postcode={00133}, 
            %state={},
            country={Italy}}

%% Abstract
\begin{abstract}
%Abstract text.
Heavy and light particles are commonly found in many natural phenomena and industrial processes, such as suspensions of bubbles, dust, and droplets in incompressible turbulent flows. Based on a recent machine learning approach using a diffusion model that successfully generated single tracer trajectories in three-dimensional turbulence and passed most statistical benchmarks across time scales, we extend this model to include heavy and light particles. Given the particle type - tracer, light, or heavy - the model can generate synthetic, realistic trajectories with correct fat-tail distributions for acceleration, anomalous power laws, and scale dependent local slope properties. % In addition, the model shows strong generalizability for extreme events \LB{we do not have more statistics than DNS, why we say this?}, generating high-intensity and rare events not present in the training data. 
This work paves the way for future exploration of the use of diffusion models to produce high-quality synthetic datasets for different flow configurations, potentially allowing interpolation between different setups and adaptation to new conditions.
\end{abstract}



%% Keywords
\begin{keyword}



\end{keyword}

\end{frontmatter}



%\section{Example Section}
%\label{sec1}
%% Labels are used to cross-reference an item using \ref command.

%Section text. See Subsection \ref{subsec1}.

\section{Introduction}

The Lagrangian description of turbulence involves tracking the information acquired by individual particles carried by the flow, and provides crucial insights into the physics underlying many natural phenomena and applied processes, such as cloud formation, industrial mixing, pollutant dispersion and quantum fluids \citep{la2001fluid, falkovich2002acceleration, yeung2002lagrangian, post2002modeling, shaw2003particle, toschi2009lagrangian, xia2013lagrangian, bentkamp2019persistent, laussy2023shining}. Lagrangian particles can convolve spatial and temporal information over an extensive range of scales. The time scale separation in a turbulent flow is given by $\tau_L/\tau_\eta \propto R_\lambda$, where $\tau_L$ is the largest energy-injection time scale and $\tau_\eta$ is the smallest Kolmogorov time scale. The Taylor microscale Reynolds number, $R_\lambda$, ranges from a few thousand in the laboratory to millions and beyond in realistic flows \citep{frisch1995turbulence}. Another intriguing feature of Lagrangian turbulence is the strong intermittency of intense fluctuations associated with small-scale vortical structures \citep{mordant2004experimentala, biferale2005particle}, which can easily lead to acceleration events in excess of 50-60 standard deviations in table-top laboratory flows \citep{voth2001silicon, mordant2004experimentalb}. Compared to tracers, which exactly follow the local flow, the situation becomes more complicated when the inertial effects of particles are combined with intermittent turbulent properties, which are important in facilitating droplet collisions and the formation of large droplets in clouds \citep{falkovich2002acceleration, kostinski2005fluctuations}. Inertial particles depart from fluid streamlines, resulting in a non-uniform spatial distribution, a phenomenon known as preferential concentration \citep{toschi2009lagrangian}. Light particles tend to accumulate in vortical structures, while heavy particles are expelled from these regions \citep{maxey1983equation, balkovsky2001intermittent, bec2003fractal, chen2006turbulent}. 
Stochastic modeling of Lagrangian tracer properties is exceptionally challenging due to multi-time dynamics, such as small-scale trapping within vortices for periods exceeding the local eddy turnover time \citep{wilson1996review, lamorgese2007conditionally, minier2014guidelines, biferale2005particle, toschi2005acceleration}. Typical modeling approaches involve proposing a random process in time for the velocity to capture the dynamics at the two spectrum extremes, $\tau_L$ and $\tau_\eta$ \citep{sawford1991reynolds, pope2011simple}. Recently, these models have been generalized to be infinitely differentiable with intermittent scaling properties by \citet{viggiano2020modelling}. Multifractal and/or multiplicative models have been used to provide a possible analytical framework, and they can reproduce some non-trivial features of turbulent statistics \citep{biferale1998mimicking, arneodo1998random, chevillard2019skewed, sinhuber2021multi, zamansky2022acceleration, lubke2023stochastic}. Furthermore, stochastic models for the generation of heavy and light particles are even more problematic, having to integrate multi-scale properties and preferential concentration \citep{friedrich2022single}.

To generate turbulent data with the correct multiscale statistics across the full range of dynamics encountered in real turbulent environments, data-driven machine learning methods have been employed due to their powerful expressive capabilities. Generative models, which learn from the underlying distribution of large amounts of training data, are particularly suitable for this task~\citep{buzzicotti2023data}. A notable example is the Generative Adversarial Network (GAN), which has been shown to effectively capture multiscale turbulent properties in the Eulerian framework \citep{buzzicotti2021reconstruction, yu2022three, li2023multi, li2023generative}. \citet{granero2024neural} utilized a U-net optimized with carefully designed loss based on multiscale properties to generate one-dimensional stochastic fields. Specifically, in our previous work \citep{li2024synthetic}, we employed a diffusion model (DM) to generate Lagrangian tracers with accurate properties, spanning from large forcing scales, through the intermittent inertial range, to the coupled regime between inertial and dissipative scales \citep{arneodo2008universal}.

Given the previous success of DM in generating tracers with correct statistical properties across time scales, and its surprising ability to generate high-intensity rare events with realistic statistics, we now question the generalizability of the model to different particles properties, i.e. in the case where inertial effects are not negligible. Due to centrifugal/centripetal effects, it is known that heavy particles tend to experience smoother viscous fluctuations, while light particles enhance them \citep{cencini2006dynamics, bec2006effects, benzi2015homogeneous}, making the problem a very important quantitative benchmark for data-driven, equations-blind tools. Specifically, here we show that DMs are able to conditionally generate multiscale Lagrangian trajectories for inertial (heavy/light) particles and tracers at moderate/high Reynolds numbers with unprecedented quantitative agreement with the ground-truth numerical data used for training. This is a step forward towards building a stochastic multiscale model for inertial particles for different Stokes numbers $St$ and added mass coefficients $\beta$ (see next section).

\section{Materials and Methods}

\subsection{Simulations for Lagrangian Particles}

To generate a dataset of Lagrangian particles, we first performed direct numerical simulations (DNS) of the incompressible Navier-Stokes equations following the approach described in ~\citep{biferale2023turb}:
\begin{equation}
\begin{cases}
    \partial_t\bm{u}+\bm{u}\cdot\nabla\bm{u}=-\nabla p+\nu\Delta\bm{u}+\bm{F} \\
    \nabla\cdot\bm{u}=0
\end{cases}
.
\end{equation}
Here $\bm{u}$ represents the Eulerian velocity field, $\nu$ the viscosity and $\bm{F}$ the large-scale isotropic forcing. We used a standard pseudo-spectral approach, fully dealiased with the two-thirds rule, within a cubic periodic domain with a resolution of $1024^3$. The resulting Taylor microscale Reynolds number was $R_\lambda\simeq310$. Details of the simulation can be found in \citep{li2024synthetic}.

Once a statistically stationary state was reached for the underlying Eulerian flow, we seeded the flow with particles. The particles were passively advected, assumed to be sufficiently dilute to neglect collisions and not to react back on the flow. The motion of a small spherical particle with radius $a$ and density $\rho_p$ suspended in the fluid with density $\rho_f$ and velocity $\bm{u}$ can be approximated as \citep{maxey1983equation, biferale2009statistics}:
\begin{equation}
	\dot{\bm{X}}(t)=\bm{V}(t),
\end{equation}
\begin{equation}
	\dot{\bm{V}}(t)= \beta D_t \bm{u}(\bm{X}(t),t)+\frac{1}{\tau_p}\left(\bm{u}(\bm{X}(t),t)-\bm{V}(t)\right),
\end{equation}
where, $\bm{X}(t)$ and $\bm{V}(t)$ are respectively the particle position and velocity, $\beta=3\rho_f/(\rho_f+2\rho_p)$ is the density ratio between the fluid and the particle, $\tau_p = a^2/(3\beta \nu)$ is the particle response time, whose ratio with the Kolmogorov time scale $\tau_\eta$ defines the particle Stokes number, $St = \tau_p/\tau_\eta$.

For the numerical integration of Lagrangian particles, we used a sixth-order B-spline interpolation scheme to obtain the fluid velocity at the particle positions and a second-order Adams-Bashforth time marching scheme for time integration \citep{van2012efficiency}. We tracked $N_p=327680$ trajectories for each type of particle: heavy ($\beta = 0.01$), tracer ($\beta=1$) and light ($\beta = 2.5$), over a total time of $T\simeq1.3\tau_L\simeq200\tau_\eta$. Both heavy and light particles are integrated with $\tau_p =0.02$ resulting in a $St= 0.87$.
Lagrangian information was recorded every $dt_{s}\simeq 0.1\tau_\eta$, resulting in each trajectory consisting of $K=2000$ points \citep{calascibetta2023optimal, li2024synthetic}.
%Heavy
%beta - tau_St - Stokes (where tau_eta=0.023): 
%0.01 - 0.02  - 0.87 
%Light
%2.5 - 0.02  - 0.87

\subsection{Diffusion Models for Conditional Generation}

\begin{figure}
    \centering
	\includegraphics[width=\textwidth]{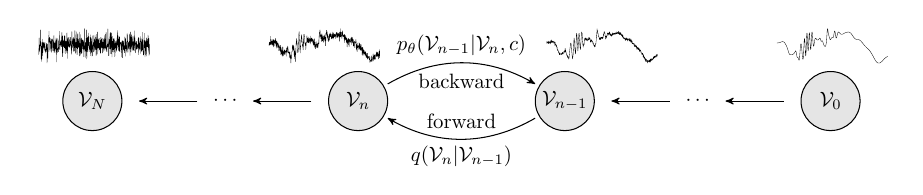}
	\caption{Illustration of the forward and backward diffusion Markov processes. The forward process (right to left) introduces noise progressively over $N$ steps. In contrast, the backward process (left to right), implemented by a neural network, generates the trajectory step by step starting from pure Gaussian noise.}\label{DM_workflow}
\end{figure}

In this section we introduce the DMs used in this work to generate Lagrangian trajectories of different particles. The DM framework consists of two main processes: the forward and the backward process. The forward process operates as a Markov chain, incrementally adding Gaussian noise to the training data until the original signal is reduced to pure noise. In contrast, as shown in Fig. \ref{DM_workflow}, the backward process starts with pure Gaussian noise and uses a learned neural network to gradually denoise and generate information, eventually producing realistic trajectory samples.
In our notation we represent each trajectory as $\CV=\{V_i(t_k)|t_k\in[0,T];i=x,y,z\}$, where $k=1,\dots,K$ are the discretized sampling times of each trajectory. The distribution of the ground-truth trajectories derived from the DNS is denoted by $q(\CV|c)$, where $c$ indicates the type of particles: tracers, heavy particles or light particles. The {\it forward} diffusion process consists of $N$ Markovian noising steps, starting from any of the trajectories generated by the DNS, $\CV_0=\CV$. Each step, $n=1,\dots,N$, is defined as
\begin{equation}
    q(\CV_n|\CV_{n-1})\to\CV_n\sim\mathcal{N}(\sqrt{1-\beta_n}\CV_{n-1},\beta_n\bm{I}),
\end{equation}
which means that $\CV_n$ samples from a Gaussian distribution with mean $\sqrt{1-\beta_n}\CV_{n-1}$ and variance $\beta_n\bm{I}$. We can formally express the forward process as
\begin{equation}
    q(\CV_{1:N}|\CV_0)\coloneqq\prod_{n=1}^{N}q(\CV_n|\CV_{n-1}),
\end{equation}
where the notation $\CV_{1:N}$ denotes the entire sequence of noisy trajectories $\CV_{1},\CV_{2},\dots,\CV_{N}$ obtained from a specific $\CV_{0}$ taken from the training set. The variance schedule $\beta_1,\ldots,\beta_N$ is predefined, with a large $N$ to allow a continuous transition to the pure Gaussian state, $\CV_N\sim\mathcal{N}(0,\bm{I})$. Further details of the variance schedule can be found in~\ref{app1}.

\begin{figure}
    \centering
 	\includegraphics[width=0.9\textwidth]{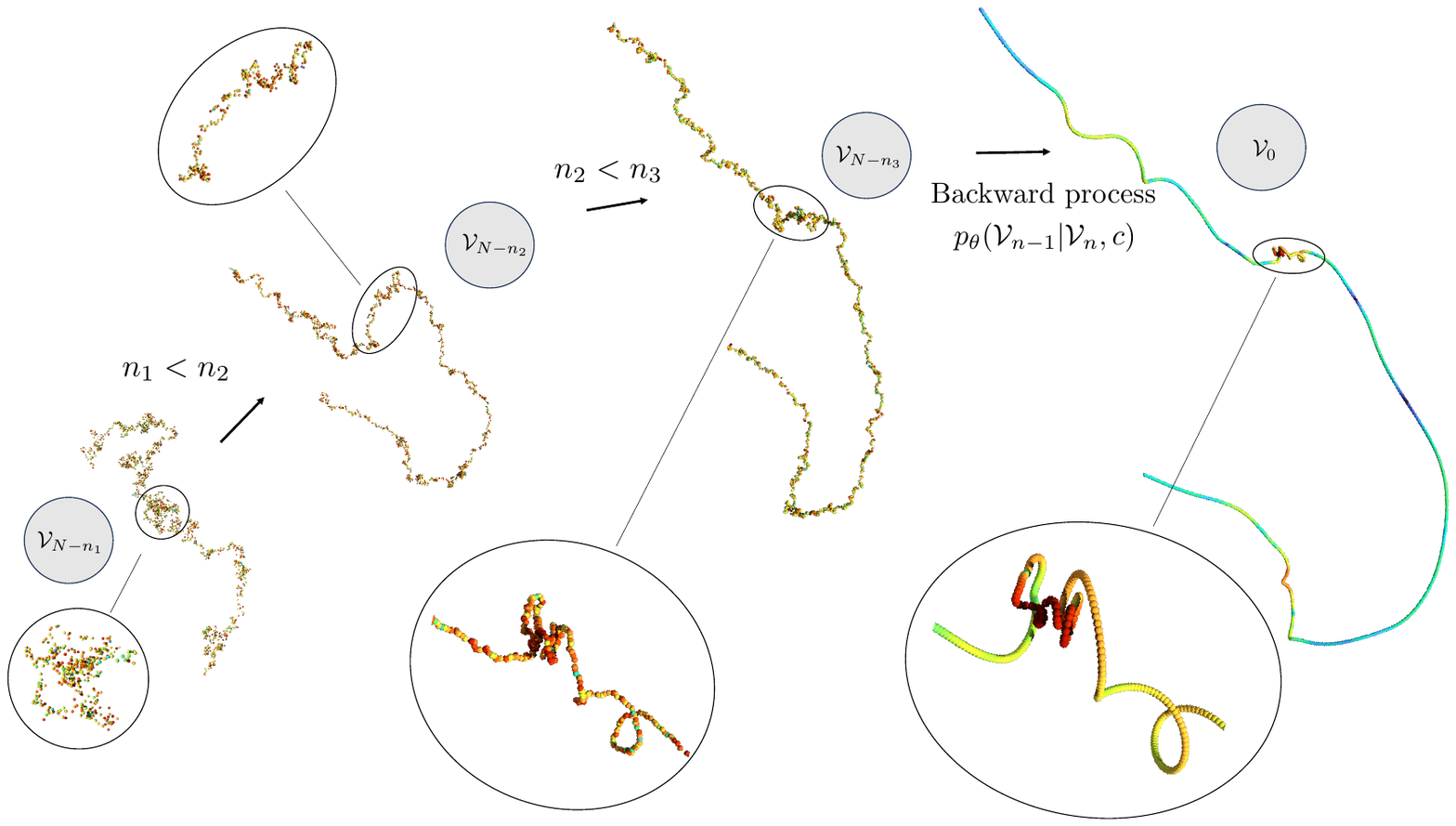}
 	\caption{Visualization of a typical tracer trajectory generation process at four different backward steps (from left to right). At each step of the generation process, the inset shows a zoomed view of the region in which a small-scale vortex structure is being generated.}\label{traj_generation}
\end{figure}

The {\it backward} process reverses the above procedure using a neural network to provide $p_\theta(\CV_{n-1}|\CV_n,c)$ for each step. Here the network uses the particle type $c$ as an additional input to condition the generation on the specific inertial properties of the trajectories we want to generate. Details of the network architecture are given in \ref{app1}. Therefore, starting with Gaussian noise drawn from $p(\CV_N) = \mathcal{N}(\bm{0},\bm{I})$, it is possible to conditionally generate new trajectories based on the desired type of particles with
\begin{equation}
    p_\theta(\CV_{0:N}|c)=p(\CV_N)\prod_{n=1}^{N}p_\theta(\CV_{n-1}|\CV_{n},c).
\end{equation}
In the continuous diffusion limit, achieved by our choice of variance schedule and number of diffusion steps, the backward step $p_\theta(\CV_{n-1}|\CV_{n},c)$ retains the same Gaussian functional form as the forward step. Therefore, the neural network is designed to predict the mean $\mu_{\theta}(\CV_n,n,c)$ and standard deviation $\Sigma_{\theta}(\CV_n,n,c)$ of the transition probability \citep{feller2015retracted, sohl2015deep}:
\begin{equation}\label{equ:backward}
    p_{\theta}(\CV_{n-1}|\CV_n,c)\to\CV_{n-1}\sim\mathcal{N}(\mu_{\theta}(\CV_n,n,c),\Sigma_{\theta}(\CV_n,n,c)).
\end{equation}
The neural network is trained to minimize an upper bound of the negative log-likelihood,
\begin{equation}\label{equ:nll}
    \mathbb{E}_{q(\CV_0|c)}[-\log(p_\theta(\CV_0|c))].
\end{equation}
A detailed derivation of the loss function can be found in \ref{app2}. In Fig.~\ref{traj_generation} we illustrate an example of a tracer trajectory that is being gradually generated along the backward process. We can see that the backward diffusion starts to reconstruct the large-scale features of the particular trajectory in the early steps, up to the generation of the small-scale intense fluctuations and the smooth regions in the final steps.

In this work we have considered two different types of DM, specifically we call DM-1c the diffusion model that is trained to generate a single velocity component along the particle trajectory, while we will call DM-3c the model that is trained to generate the three particle velocity components simultaneously. From the generation of the latter, it is possible to reconstruct the three-dimensional structure of the Lagrangian trajectory by time integration of the particle velocity.

\section{Results}
\begin{figure}
	\includegraphics[width=1.\textwidth]{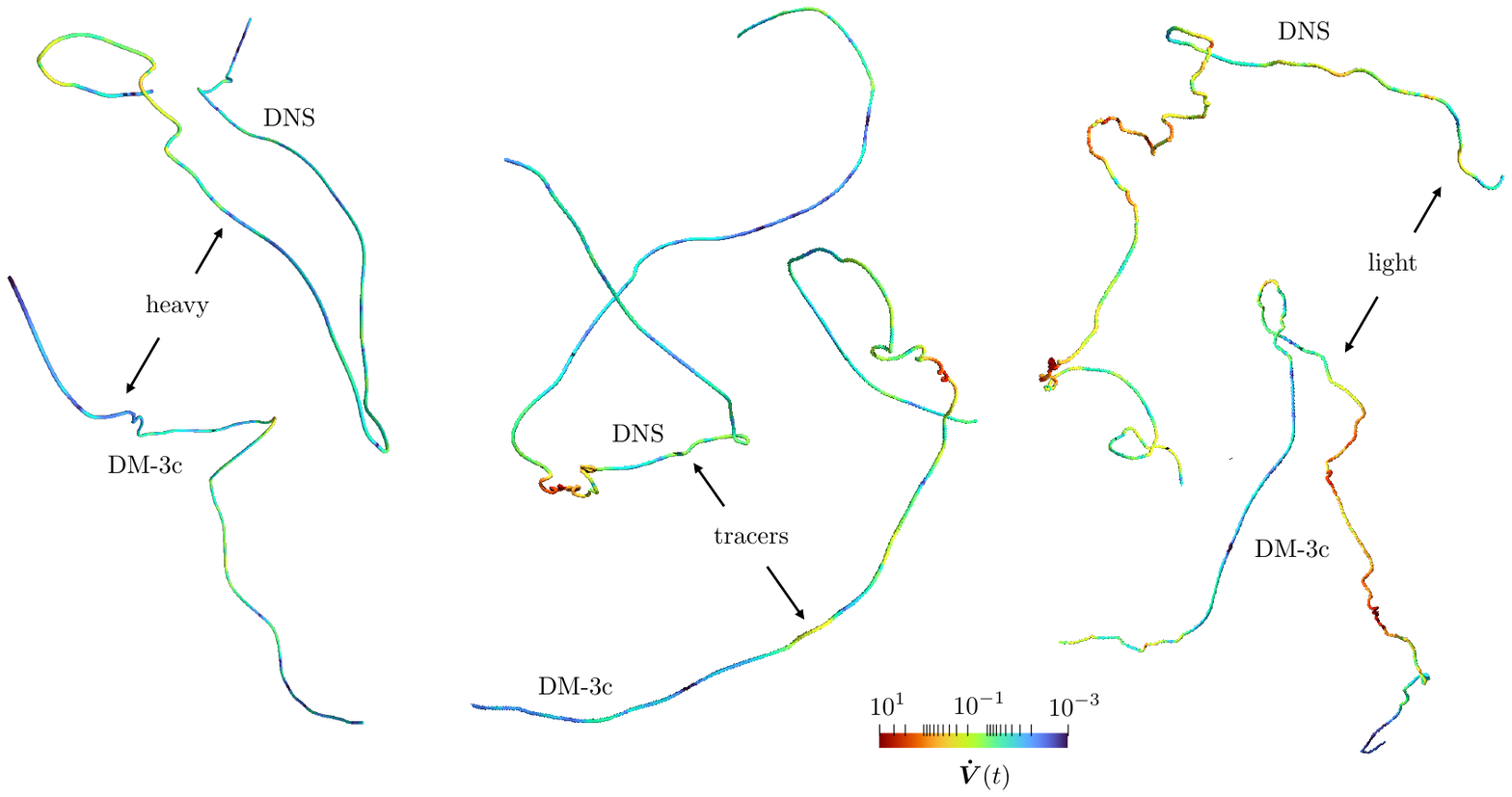}
	\caption{Examples of 3D trajectories generated from DNS and DM-3c from left to right respectively for heavy, tracers, and light particles. The colors are proportional to the local acceleration experienced by the particles along the trajectories, in particular, red indicates intense acceleration and blue indicates low acceleration regions.}\label{6trajectories}
\end{figure}

\begin{figure}
    \centering
	\includegraphics[width=1.\textwidth]{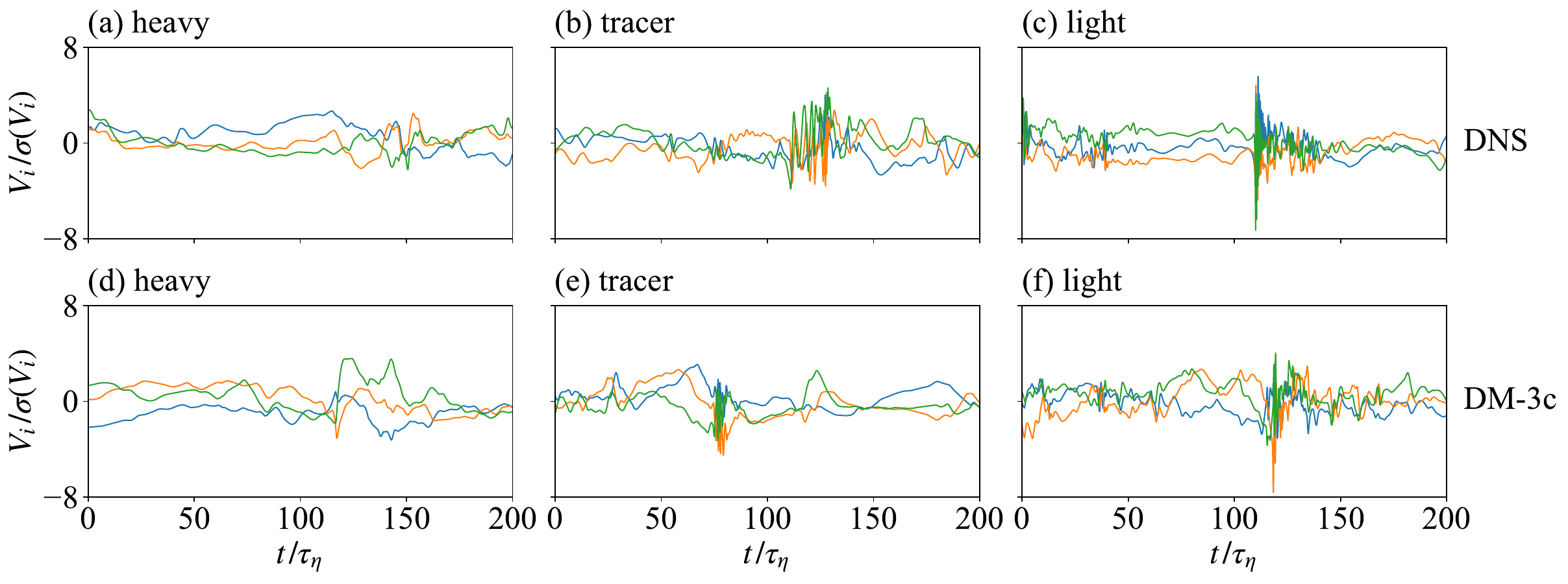}
	\caption{Examples of different velocity components $i=x,y,z$ normalized by the standard deviation $\sigma$ as a function of time for trajectories from DNS (top) and DM-3c (bottom). (a)(d) heavy particles, (b)(e) tracers, (c)(f) light particles.}\label{vel_comps}
\end{figure}

\begin{figure}
    \includegraphics[width=\textwidth]{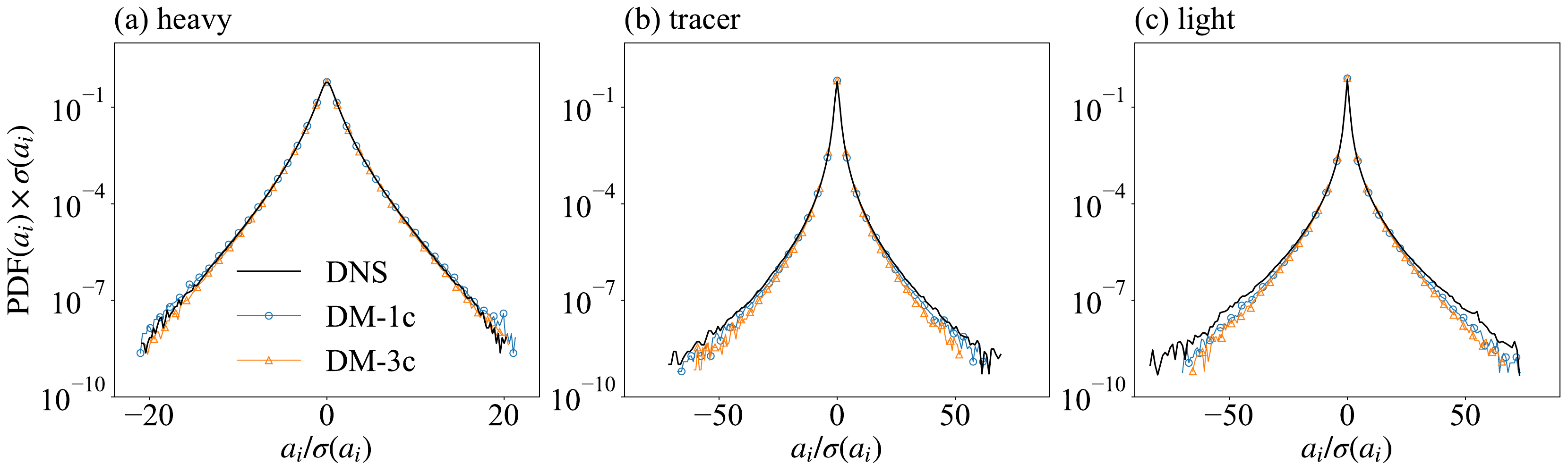}
	\caption{Standardized PDFs of one generic component of the acceleration, $a_i$, for ground-truth DNS data (black line) and synthetically generated data from DM-1c (blue line with circles) and DM-3c (orange line with triangles) for (a) heavy particles, (b) tracers, and (c) light particles. The statistics for DMs are based on the same amount of data as those for DNS.}\label{pdf_a} 
\end{figure}

\begin{figure}
    \includegraphics[width=\textwidth]{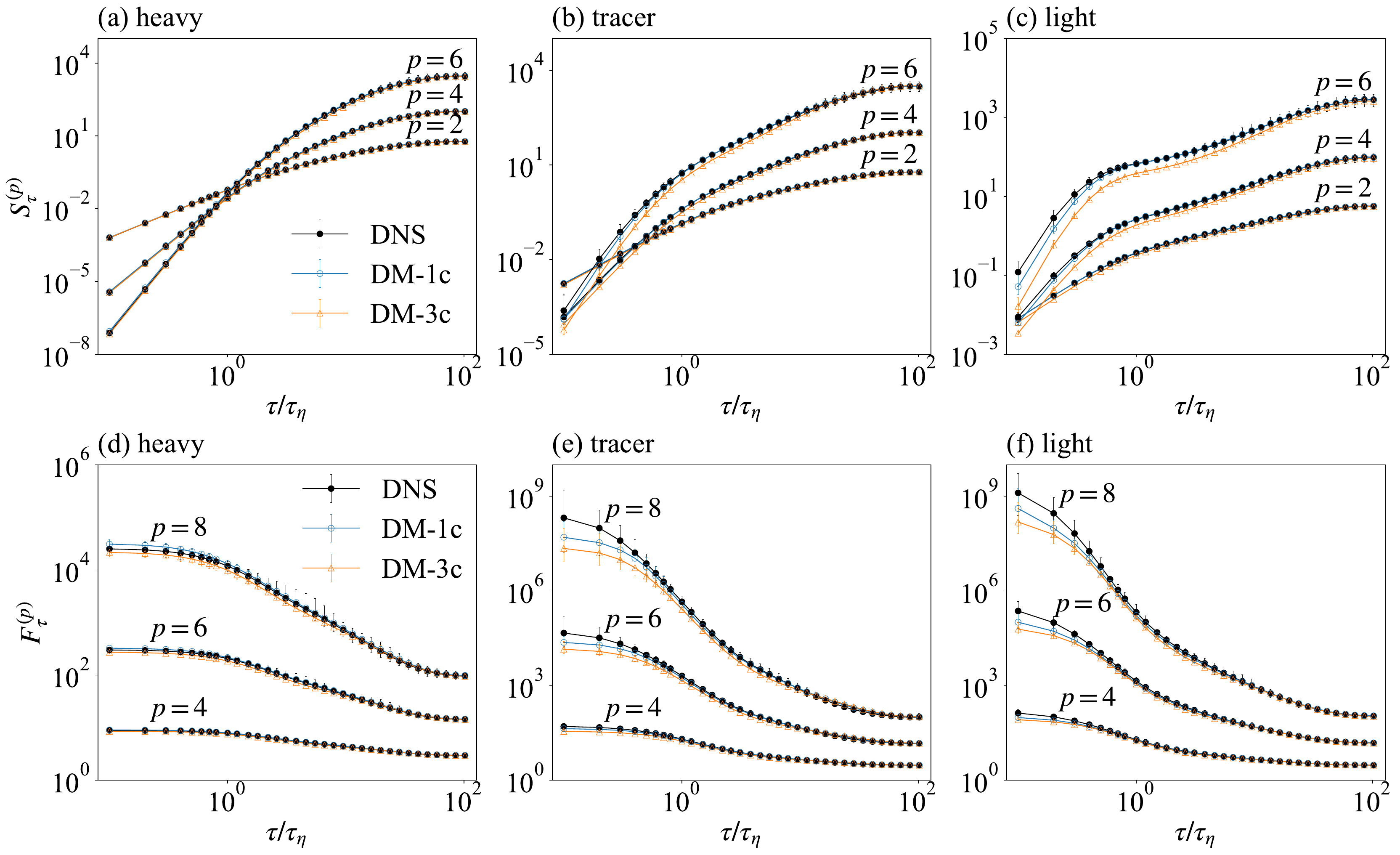}
	\caption{Log-log plots comparing Lagrangian structure functions, $S^{(p)}_\tau$ for $p=2,4,6$, and generalized flatness, $F^{(p)}_\tau$ for $p=4,6,8$, between DNS and DMs for different particle types: (a)(d) heavy particles, (b)(e) tracers, and (c)(f) light particles. The color scheme and symbols are organized as in Fig.~\ref{pdf_a}. The error bars indicate the range of values obtained for each measure by dividing the dataset used for the statistics into ten different independent batches per velocity component. This resulted in 10 batches for DM-1c and 30 batches for DNS and DM-3c.}\label{Sp}
\end{figure}

As a first result, in Fig.~\ref{6trajectories} we visually compare three-dimensional (3D) trajectories obtained from DNS with those generated by DM-3c for the three particle types considered in this work, heavy, tracer, and light. This first result is useful to show qualitatively that DM can reproduce the complex topological-vortical structures expected in the real trajectories with different inertia. From this figure, we can see that light particles, in both the DNS and DM examples, experience intense acceleration events (red-colored regions) much more frequently than the tracers, while heavy particles have a much smoother dynamic, reflecting DM's ability to correctly model the particle nature of being trapped in or escaping vortex filaments.
Figure \ref{vel_comps} shows the three velocity components as a function of time for typical trajectories of different particles obtained from DNS and DM-3c. This comparison further demonstrates the consistency between DNS and DM for particles with different properties. The increasingly obvious and intense vortex-trapping events from heavy to tracer to light particles reflect their sampling of different regions of the turbulent field.
\\
In order to have a first comparison over the whole generated and the training dataset, in Fig. \ref{pdf_a} we show the probability density function (PDF) of a generic component of the acceleration along the particle trajectories. The instantaneous particle acceleration is calculated as $a_i(t)=\lim_{\tau\to0}\delta_\tau V_i/\tau$, approximated with a time resolution of $0.1\tau_{\eta}$. We can see that there is for all cases a very close agreement between the ground-truth DNS distributions and those from DMs over the whole range of fluctuations, and up to the extreme fluctuations, $60-70$ times the standard deviation, observed for tracers and light particles.
For a more quantitative comparison, we now study the statistical properties of high-order two-point time correlations by introducing the so-called Lagrangian structure functions, defined as
\begin{equation}
S_\tau^{(p)}=\langle[V_i(t+\tau)-V_i(t)]^p\rangle,
\end{equation}
where on the l.h.s. we have removed the dependency on the component $i=x,y,z$ assuming isotropy. Fig.~\ref{Sp} (top row) shows the Lagrangian structure functions of order $p=2,4,6$ for the DNS training data, the data set generated by DM-1c, which generates individual velocity components, and DM-3c, which generates all three velocity components simultaneously. In the bottom row of the same figure, we show a comparison of the generalized flatness, 
\begin{equation}
F_\tau^{(p)}= S_\tau^{(p)} / [S_\tau^{(2)}]^{p/2} ,
\end{equation}
up to $p=8$ obtained for the same datasets discussed above. Given the sensitivity to the rare fluctuations in the data of such high-order observables and their extension over more than two decades of dynamical scales, it is remarkable how accurately the DM reproduces the correct ground truth statistics while distinguishing the different particle phenomena.

\begin{figure}
    \includegraphics[width=\textwidth]{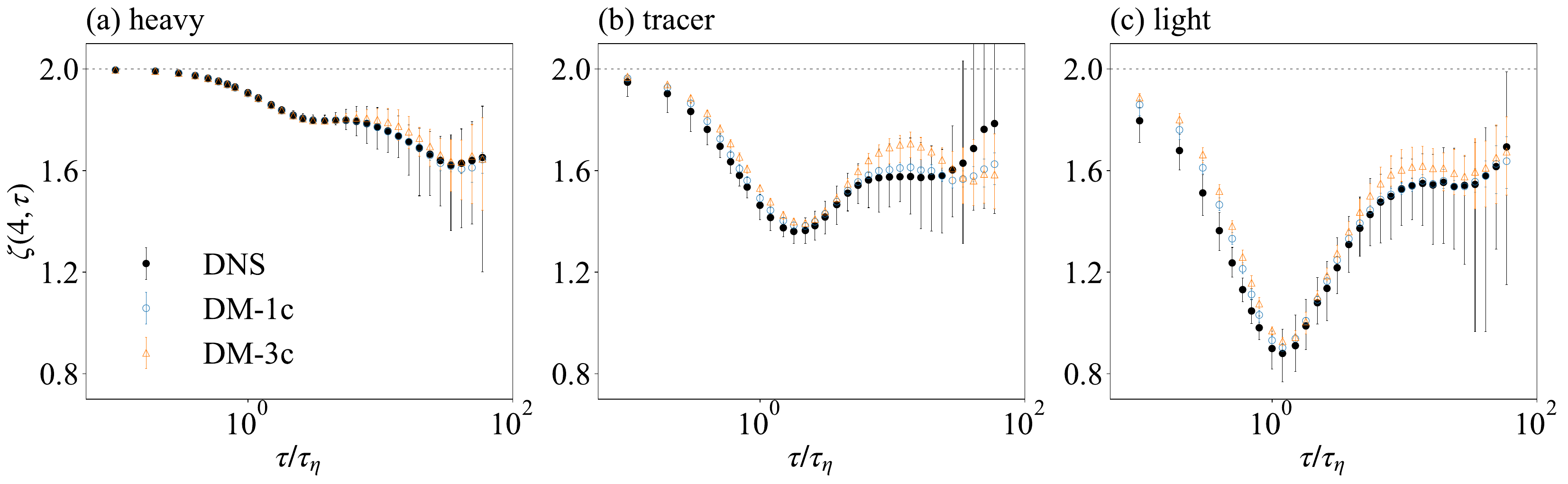}
    \caption{Comparison of 4th-order logarithmic local slope $\zeta(4,\tau)$ between the ground-truth DNS and DMs on a lin-log scale for (a) heavy particles, (b) tracers, and (c) light particles.  The dotted horizontal lines represent the non-intermittent dimensional scaling, $S^{(4)}_\tau \propto [S^{(2)}_\tau]^2$. The color scheme and symbols are organized as in Fig.~\ref{pdf_a}, while the statistics and the error bars are derived in the same way as in Fig. \ref{Sp}.}\label{local_slope}
\end{figure}

Finally, we discuss the most rigorous multiscale statistical test: comparing the scale-by-scale exponents obtained from the logarithmic derivatives of the structure functions in extended self-similarity (ESS)~\citep{arneodo2008universal}, namely computed as
\begin{equation}
    \zeta(p,\tau)=\frac{d\,\log S_\tau^{(p)}}{d\,\log S_\tau^{(2)}}.
\end{equation}
%To compute $\zeta(p,\tau)$, we divide the results of $d\log S_\tau^{(p)}/d\log\tau$ and $d\log S_\tau^{(2)}/d\log\tau$ \LB{??? I do not understand the previous statement}\TL{, which} are computed on a grid with $\tau$ intervals of 1 (from 1 to 1024) using second-order accurate central differences \TL{[TL:is above sentence clear now?]}. 
%For tracer particles, multifractal statistical models cannot accurately capture the complexity of the $\zeta(p,\tau)$ curves over all time scales from forcing to viscous regimes \citep{nelkin1990multifractal, borgas1993multifractal, chevillard2003lagrangian}. 
To our knowledge, DM is the first method to successfully generate synthetic 3D Lagrangian tracer trajectories that reproduce this observable across all time scales~\citep{li2024synthetic}. In Fig.~\ref{local_slope} we show the ESS local exponent for $p=4$, again comparing DNS, DM-1c and DM-3c for the three particle types. These results allow us to conclude that DMs can correctly capture the multiscale properties of the structure-function scaling exponents even in the presence of different inertial properties. In particular, we can see how the model is able to correctly reproduce the different vortex trapping dynamics, which is strongly enhanced for light particles and depleted for heavy ones compared to the tracers, which is reflected in the intensification of the intermittency level and the depth of the viscous bottleneck around the range $\tau\sim\tau_\eta$ while decreasing the particle inertia, from heavy to tracer to light.

\section{Conclusions}
We have generalized a data-driven diffusion model, originally successful in generating single-particle tracers, to accommodate particle with different inertia: tracers, heavy and light particles. By incorporating data from different particle inertia, the model has adapted to new conditions while maintaining its effectiveness. It reproduces most statistical benchmarks across time scales, including the fat-tail distribution for acceleration, the anomalous power law, and the increased intermittency around the dissipative scale for tracers and light particles. Note that the original model showed a strong ability to generate unseen extreme events \citep{li2024synthetic}; future work will involve collecting more statistics to check for similar capabilities in the current model.%The model also shows strong generalizability for extreme events, producing unseen high-intensity and rare events that still match realistic statistics\LB{do we have more statistics?}.

In future research, the generalizability of the DM model can be further tested by including a more diverse set of data configurations in the training process. This will allow us to evaluate the interpolation and extrapolation capabilities of the model for various physical parameters such as density ratios, Stokes numbers, and Reynolds numbers to fully explore the potential of the model. Advanced network architectures, such as transformers \citep{vaswani2017attention}, could be used to replace the current U-net to better handle the scaling capability required for larger and more diverse datasets. Our ultimate goal is to provide high-quality, high-volume synthetic datasets for downstream applications, such as inertial particle classification and data inpainting \citep{friedrich2020stochastic, li2023multi_dm, zheng2024high}, thereby avoiding the unfeasible computational or experimental effort required to generate real Lagrangian trajectories.% Data generation from DM is not only faster for the same physical parameters, but would also be capable of interpolation across different physical parameter spaces.

\section*{Acknowledgements}
We thank Mauro Sbragaglia and Roberto Benzi for useful discussions and collaborations in a early stage of this work. This work was supported by the European Research Council (ERC) under the European Union’s Horizon 2020 research and innovation programme Smart-TURB (Grant Agreement No. 882340), by the MUR-FARE project R2045J8XAW, and by  Next Generation EU, Piano Nazionale di Ripresa e Resilienza (PNRR), Missione 4 “Istruzione E Ricerca” 
Componente C2 Investimento 1.1. Fondo per il programma Nazionale di
Ricerca e Progetti di Rilevante Interesse Nazionale (PRIN) 202249Z89M.

\appendix
% \section{Example Appendix Section}


\section{DM architecture and noise schedule}
\label{app1}

\begin{figure}
    \centering
	\includegraphics[width=\textwidth]{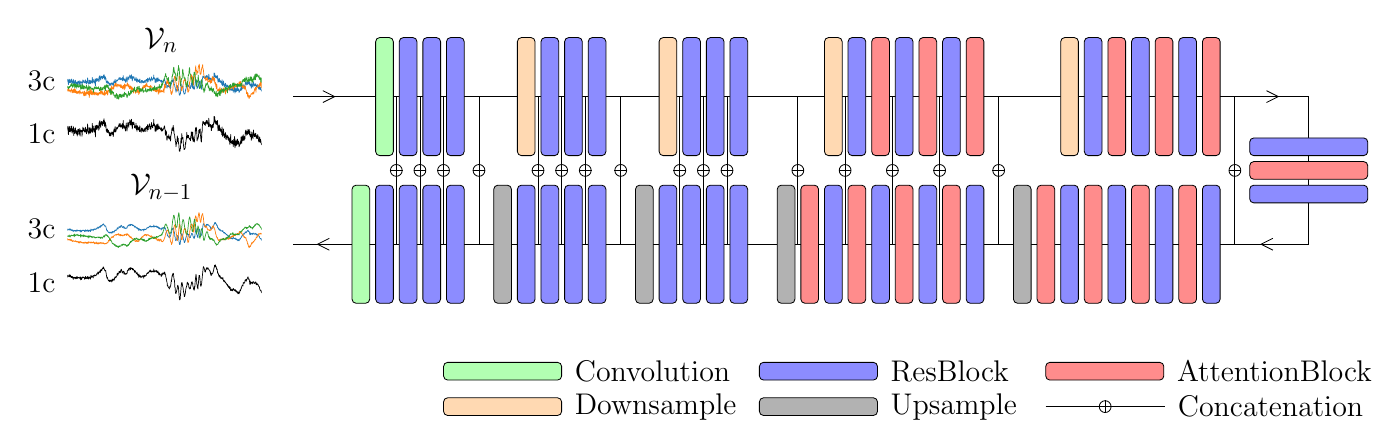}
	\caption{The U-net architecture that takes a noisy trajectory of a given particle inertia as input at step $n$ and predicts a denoised trajectory at step $n-1$.}\label{U-net}
\end{figure}

We use a U-net architecture \citep{ronneberger2015u} as the backbone of the DM, consisting of two main parts: a downsampling stack and an upsampling stack, connected by skip connections as shown in Fig. \ref{U-net}. The upsampling stack mirrors the downsampling stack, creating a symmetrical structure, with each stack performing four steps of downsampling or upsampling, respectively. This results in five stages from highest to lowest resolution (2000 to 125, each with a downsampling/upsampling rate of 2). The three residual blocks in these stages are configured with channels $[C,C,2C,3C,4C]$, with $C$ set to 128. Multi-head attention \citep{vaswani2017attention} with four heads is implemented after each residual block in the 250 and 125 resolution stages. The intermediate module connecting the encoder and decoder stacks consists of two residual blocks of $4C$ channels, sandwiching a four-head attention. The diffusion step $n$ is specified to the network using transformer sinusoidal position embedding and the particle type is specified using class embedding.

We adopted the optimal noise schedule from previous research to generate Lagrangian tracers with a total of $N=800$ diffusion steps \citep{li2024synthetic}:
\begin{equation}
    \bar{\alpha}_n=\frac{-\tanh{(7n/N-6)}+\tanh{1}}{-\tanh{(-6)}+\tanh{1}}.
\end{equation}
The variance can be obtained as $\beta_n=1-\bar{\alpha}_n/\bar{\alpha}_{n-1}$, which is clipped to be no greater than 0.999 to avoid singularities at the end of the forward diffusion.

The AdamW optimizer \citep{loshchilov2017decoupled} was used to train the model with a learning rate of $10^{-4}$ over $2.5\times10^5$ iterations for DM-1c, and $4.0\times10^5$ iterations for DM-3c. The DMs were trained with a batch size of 256 on four NVIDIA A100 GPUs for approximately 25 hours (DM-1c) and 40 hours (DM-3c). An exponential moving average (EMA) strategy with a decay rate of 0.999 was applied to the model parameters to sample new trajectories.

\section{Derivation of the Training Loss Function}
\label{app2}

We introduce an important property of the forward process: it allows closed-form sampling of $\CV_n$ at each diffusion step $n$ \citep{weng2021diffusion}:
\begin{equation}
    q(\CV_n|\CV_0)\to\CV_n\sim\mathcal{N}(\sqrt{\bar{\alpha}_n}\CV_0,(1-\bar{\alpha}_n)\bm{I}),
\end{equation}
where we define $\alpha_n\coloneqq1-\beta_n$ and $\bar{\alpha}_n\coloneqq\Pi_{i=1}^{n}\alpha_i$. In particular, given any initial trajectory $\CV_0$, its state after $n$ diffusion steps can be sampled directly as
\begin{equation}\label{equ:q_sample}
    \CV_n=\sqrt{\bar{\alpha}_n}\CV_0+\sqrt{1-\bar{\alpha}_n}\epsilon,
\end{equation}
where $\epsilon\sim\mathcal{N}(\bm{0},\bm{I})$.

We use the variational bound to optimize the negative log-likelihood in Eq. (\ref{equ:nll}):
\begin{equation}
    L\coloneqq\mathbb{E}_{q(\CV_0)}\mathbb{E}_{q(\CV_{1:N}|\CV_0)}\left[-\log\frac{p_\theta(\CV_{(0:N)})}{q(\CV_{1:N}|\CV_0)}\right]\ge\mathbb{E}_{q(\CV_0)}[-\log(p_\theta(\CV_0))].
\end{equation}
From this point on, we omit the condition on particle type $c$ for the sake of simplicity. This objective can be expressed as the sum of the Kullback–Leibler (KL) divergences, denoted as $\kl{\cdot}{\cdot}$, together with an additional entropy term \citep{sohl2015deep, ho2020denoising}:
\begin{align}
    L=\;&\mathbb{E}_{q(\CV_0)}\bigg[\underbrace{\kl{p(\CV_N|\CV_0)}{p_{\theta}(\CV_N)}}_{L_N}\nonumber\\
    &+\sum_{n>1}^N\underbrace{\kl{p(\CV_{n-1}|\CV_n,\CV_0)}{p_\theta(\CV_{n-1}|\CV_n)}}_{L_{n-1}}\underbrace{-\log p_\theta(\CV_0|\CV_1)}_{L_0}\bigg].
\end{align}
The first term, $L_N$, is ignored during training because it contains no learnable parameters, since $p_\theta(\CV_N)$ is a Gaussian distribution. The second part of the terms, $L_{n-1}$, represents the KL divergence between $p_\theta(\CV_{n-1}|\CV_n)$ and the posteriors of the forward process conditioned on $\CV_0$, which are tractable using Bayes' theorem \citep{weng2021diffusion}:
\begin{equation}\label{equ:posterior}
    p(\CV_{n-1}|\CV_n,\CV_0)\to\CV_{n-1}\sim\mathcal{N}(\tilde{\mu}(\CV_n,\CV_0),\tilde{\beta}_n\bm{I}),
\end{equation}
where
\begin{equation}\label{equ:posterior_mean}
    \tilde{\mu}_n(\CV_n,\CV_0)\coloneqq\frac{\sqrt{\bar{\alpha}_{n-1}}\beta_n}{1-\bar{\alpha}_n}\CV_0+\frac{\sqrt{\alpha_n}(1-\bar{\alpha}_{n-1})}{1-\bar{\alpha}_n}\CV_n
\end{equation}
and
\begin{equation}
    \tilde{\beta}_n\coloneqq\frac{1-\bar{\alpha}_{n-1}}{1-\bar{\alpha}_n}\beta_n.
\end{equation}
The KL divergence between the two Gaussians in Eqs. (\ref{equ:backward}) and (\ref{equ:posterior}) can be expressed as
\begin{equation}\label{equ:Lnm1_mu}
    L_{n-1}=\mathbb{E}_{q(\CV_0)}\left[\frac{1}{2\sigma_n^2}\|\tilde{\mu}_n(\CV_n,\CV_0)-\mu_\theta(\CV_n,n)\|^2\right],
\end{equation}
given the constant variance $\Sigma_\theta=\sigma_n^2\bm{I}$, where $\sigma_n^2$ can be either $\beta_n$ or $\tilde{\beta}_n$ as discussed in \citep{ho2020denoising} and we use the former in this work. It can be shown that the term $L_0$ takes the same form as in Eq. (\ref{equ:Lnm1_mu}) due to the Gaussian form of $p_\theta(\CV_0|\CV_1)$ in Eq. (\ref{equ:backward}).

We aim to train $\mu_\theta(\CV_n,n)$ to predict $\tilde{\mu}(\CV_n,\CV_0)$, which is given by
\begin{equation}
    \tilde{\mu}(\CV_n,\CV_0)=\frac{1}{\sqrt{\alpha_n}}\left(\CV_n-\frac{\beta_n}{\sqrt{1-\bar{\alpha}_n}}\epsilon\right),
\end{equation}
by substituting Eq. (\ref{equ:q_sample}) into Eq. (\ref{equ:posterior_mean}). Therefore, given that $\CV_n$ is available as input to the model, we can reparameterize to make the network predict the Gaussian noise term $\epsilon$, and the predicted mean is
\begin{equation}
    \mu_\theta(\CV_n,n)=\frac{1}{\sqrt{\alpha_n}}\left(\CV_n-\frac{\beta_n}{\sqrt{1-\bar{\alpha}_n}}\epsilon_\theta(\CV_n,n)\right),
\end{equation}
where $\epsilon_\theta$ is the predicted cumulative noise at step $n$. This reparameterization transforms Eq. (\ref{equ:Lnm1_mu}) into
\begin{equation}
    L_{n-1}=\mathbb{E}_{q(\CV_0),\,\epsilon}\left[\frac{\beta_n^2}{2\sigma_n^2\alpha_n(1-\bar{\alpha}_n)}\|\epsilon-\epsilon_\theta\left(\CV_n(\CV_0,\epsilon),n\right)\|^2\right].
\end{equation}
We further ignore the weighting term and optimize a simplified version of the variational bound:
\begin{equation}\label{equ:L_simple}
    L_\mathrm{simple}=\mathbb{E}_{n,\,q(\CV_0),\,\epsilon}\left[\|\epsilon-\epsilon_\theta\left(\CV_n(\CV_0,\epsilon),n\right)\|^2\right],
\end{equation}
where $n$ is sampled uniformly from $1$ to $N$. In practice, this method improves the sample quality and simplifies the implementation \citep{ho2020denoising}.

%%       \citet{<label>} ==> Lamport (1994)
%%       \citep{<label>} ==> (Lamport, 1994)
%%
%Example citation, See \citet{lamport94}.


\bibliographystyle{elsarticle-harv} 
\bibliography{ref}



%\begin{thebibliography}{00}
%
%%% For authoryear reference style
%%% \bibitem[Author(year)]{label}
%%% Text of bibliographic item
%
%\bibitem[Lamport(1994)]{lamport94}
%  \textit{\LaTeX: a document preparation system},
%  Addison Wesley, Massachusetts,
%\end{thebibliography}
\end{document}